\documentclass[11pt,preprint]{aastex}
\received{}
\accepted{}

\slugcomment{Accepted by ApJ}
\lefthead{Wilkes et al.}
\righthead{3C 270.1 with \chandra}

\usepackage{graphics}
\usepackage{graphicx}
\usepackage{bm}
\usepackage{rotating}
\usepackage{natbib}

\usepackage{times,mathptm,graphicx}

\def\q24 {q$_{24}$}

\def\deg {$^{\rm o}$}

\def\nh {${\rm N_H}$}
\def\lax    {${_<\atop^{\sim}}$}

\def\chandra {{\it Chandra}}

\def\ang    {\rm \AA\/}

\begin{document}

\title{\chandra\ X-ray Observations of the redshift 1.53, radio-loud
  quasar: 3C\,270.1.}
\author{Belinda J. Wilkes$^1$, Dharam V. Lal$^1$, D.M.Worrall$^2$, 
Mark Birkinshaw$^2$, Martin Haas$^3$, S.P. Willner$^1$, 
Robert Antonucci$^4$, M.L.N. Ashby$^1$, Mark Avara$^5$, 
Peter Barthel$^6$, Rolf Chini$^{3,12}$, G.G Fazio$^1$, Martin Hardcastle $^7$, 
Charles Lawrence$^8$, Christian Leipski$^9$, Patrick Ogle$^{10}$, Bernhard
Schulz$^{11}$
}
\affil{1: Harvard-Smithsonian Center for Astrophysics, Cambridge, MA 02138}
\affil{2: HH Wills Physics Laboratory, University of Bristol, UK}
\affil{3: Astronomisches Institut, Ruhr-University, Bochum, Germany}
\affil{4: Department of Physics, University of California, Santa
  Barbara, CA 93106}
\affil{5: Department of Astronomy, University of Maryland, College
  Park, MD 20742-2421}
\affil{6: Kapteyn Astronomical Institute, University of Groningen, 
The Netherlands}
\affil{7: School of Physics and Astronomy, University of Hertfordshire, England}
\affil{8: JPL, Pasadena, CA 91109}
\affil{9: MPIA Heidelberg, Germany}
\affil{10: Spitzer Science Center, Caltech, Pasadena, CA 91125}
\affil{11: IPAC, Caltech, 770 S. Wilson Ave., Pasadena,  CA 91125}
\affil{12: Instituto de Astronom\'{i}a, Universidad Cat\'{o}lica del Norte,
Antofagasta, Chile}

\begin{abstract}
\chandra\ X-ray observations of the high redshift ($z$=1.532)
radio-loud quasar 3C\,270.1  in  2008 February
show the nucleus to have a
power-law spectrum, $\Gamma = 1.66 \pm 0.08$,
typical of a radio-loud quasar, and a marginally-detected
Fe K$\alpha$ emission line. 
The data also reveal extended X-ray emission, about half of
which is associated with the radio emission from this
source. The southern emission is
co-spatial with the radio lobe and peaks at the position of the
double radio hotspot. Modeling this hotspot including
Spitzer upper limits rules out synchrotron
emission from a single power-law population of electrons, 
favoring inverse-Compton emission  with a field of $\sim$11\,nT,
roughly a third of the equipartition value. The northern
emission is concentrated close to the location of 
a 40\deg\ bend where the radio jet is presumed to 
encounter external material. 
It can be explained by inverse Compton emission
involving Cosmic Microwave Background photons with a field
of $\sim$3\,nT, roughly a factor of nine below the equipartition value.
The remaining, more diffuse X-ray emission is harder (HR=$-0.09\pm$0.22).
With only 22.8$\pm$5.6 counts, the spectral form cannot be
constrained. Assuming thermal emission with a temperature
of 4 keV yields an estimate for the luminosity of
1.8$\times 10^{44}$ erg s$^{-1}$, consistent with the
luminosity-temperature relation of
lower-redshift clusters. However deeper 
\chandra\ X-ray observations are required 
to delineate the spatial distribution, and better constrain the spectrum 
of the diffuse emission to verify that we have 
detected X-ray emission from a high-redshift cluster.
\end{abstract}


\section{Introduction}

With the aim of quantifying the orientation-dependence of
the observed properties of quasars, 
we have embarked on multi-wavelength observations of
high-redshift (1$\leq z \leq $ 2) radio sources with
\chandra , {\it Spitzer, Herschel} and ground-based observatories. 
Given the known orientation dependence of the emission from quasars
and active galactic nuclei (AGN) 
and the resulting strong selection bias against obscured/edge-on
sources, isotropic, low-frequency 
radio emission provides a rare, unbiased 
view of the population based on optically thin emission far from the
nucleus ({\it i.e.} lobe-selection). 
We chose high-redshift 3CR (selected at 178 MHz)
sources to ensure the sample is largely 
unbiased and that it comprises powerful radio galaxies and quasars.
Studies of high-redshift, radio-loud quasars also facilitate
more lines of investigation: X-ray emission related to
radio structure at high-redshift, and thus the interaction of the quasar
with its environment, and searches for high-redshift
clusters of galaxies.

X-ray emission is often 
observed from radio-emitting hotspots and lobes in quasars.
It is generally interpreted \citep{2002ApJ...565..244H, 
2009A&ARv..17....1W}
in terms of direct synchrotron emission from high-energy electrons,
inverse-Compton (iC) emission due to up-scattering of radio photons
within the radio hotspots 
(synchrotron self-Compton (SSC), \citet{2004ApJ...612..729H}),  or
inverse-Compton (iC) up-scattering of external radio photons in the extended
lobes, most likely from the Cosmic Microwave Background (iC/CMB, 
\citet{2005ApJ...626..733C}). 
The latter will be more luminous for larger radio-emitting
structures and at higher redshift due to the higher energy density of the
CMB.

Luminous high-redshift radio sources occur in massive galaxies and
so, according to the hierarchical paradigm, form at peaks in the dark
matter density. Thus they are beacons for high density regions
in the early universe and for high-redshift clusters
and groups, very few of which are yet known. 
Finding clusters at these high redshifts is key to
the study of both cluster and galaxy formation and will provide
critical constraints on theoretical
models for cluster and galaxy evolution
\citep{2011MNRAS.tmp...73A,2010ApJ...718..133H,2007A&A...461..823V} and the
mass distribution of dark matter halos \citep{2007MNRAS.374.1303C}. 

The radio-loud quasar 3C\,270.1 ($z$=1.532) has the double-lobed radio
structure characteristic of FR-II \citep{1974MNRAS.167P..31F}
radio sources and the strong, broad
emission lines of a type 1 quasar. 
Optical and
infrared (IR) data show an excess of galaxies within $\sim 640$ kpc
(1.\arcmin3) of the quasar suggesting a surrounding cluster of galaxies
\citep{2009ApJ...695..724H}. The sky distribution of the cluster
galaxy candidates forms a loose concentration
centered $\sim 20 \arcsec$ east of the quasar, although the
centroid is not well constrained by the current data.

This paper reports \chandra\ observations of
the X-ray emission from the vicinity of 3C\,270.1. This includes 
unresolved emission from the quasar itself and extended X-ray
emission within $\sim 10 \arcsec$ of the core. 
The extended X-rays have two components: softer emission associated with 
the radio emission north and south of the core; and harder, more
diffuse emission
which may originate in gas associated with the
cluster in which the quasar is embedded \citep{2009ApJ...695..724H}
or in iC/CMB emission from low-surface brightness or older radio
emission that is not easily detectable
\citep{2003MNRAS.341..729F}.

We assume a $\Lambda$CDM cosmology with H$_o$=71 km s$^{-1}$ Mpc$^{-1}$,
$\Omega_M=0.27$, and $\Omega_{vac}=0.73$ \citep{2011ApJS..192...16L}.

\section{Multi-wavelength Observations and Data}

3C\,270.1 was observed on-axis with the \chandra\ 
ACIS-S for 9.673 ks on 2008 February 16 (\dataset[ADS/Sa.CXO#obs/09255]
{ObsId 9255}). 
The data 
were reprocessed in 2011 January with CIAO 4.3
to take advantage of the latest calibration files and CTI (charge
transfer inefficiency) calibration
as well as sub-pixel event repositioning, which improves the spatial
resolution. The reprocessed data were used in the 
analysis presented here. 
The quasar is well-detected with $>700$ counts consistent with the
spatial distribution of a point source. There are also $>$40 excess
counts within 10\arcsec\ of the quasar but outside its' point spread
function (PSF).


Radio data from the VLA archive were reprocessed to generate
high-resolution images at 1.43, 8.46, and 14.94 GHz, to provide
comparison images at resolution similar to, or better than, that
of the \chandra\ data. All processing was done in AIPS, and
followed the usual steps of calibration, imaging, and a
CLEANing/self-calibration cycle.
The datasets used are listed in Table~\ref{tb:vla}, with the angular sizes of
the synthesized beams generated from the full datasets at uniform
sampling.

\begin{table}[h]
\caption{VLA Radio data}
\label{tb:vla}
\begin{tabular}{cccc}
\hline 
Project&Date Observed& Frequency& Full-resolution Gaussian \\
          &              & (GHz)    & beam FWHM  \\
          &              &          &   major $\times$ minor (PA) \\
          &              &               &   (arcsec$^2$) (deg) \\
\hline 
    AB796 &  1996-Nov-07 &   8.46   &  0.25 $\times$ 0.24
    (+120.5$^{\circ}$) \\ 
    AH480 &  1992-Dec-05 &   1.43   &  1.30 $\times$ 1.18
    (+35.9$^{\circ}$) \\ 
    AL073 &  1983-Oct-21 &  14.94   &  0.13 $\times$ 0.12
    (+69.4$^{\circ}$) \\ 
\hline
\end{tabular}
\end{table}


{\it Spitzer/IRAC} data for 3C\,270.1 were obtained on 
2007 June 28
and are reported by \citet{2008ApJ...688..122H,2009ApJ...695..724H}.
The on-source exposure times of 4$\times$30s in each band allow us to
detect point sources to $\sim 3~\mu$Jy (3$\sigma$) at 3.6 and 
4.5~$\mu$m and $\sim 25~\mu$Jy at 5.8 and 8~$\mu$m.

We obtained $z^\prime$ ($\lambda_{eff} \sim 9049$ \ang, 
\citet{1996AJ....111.1748F})
imaging of 3C\,270.1 in 2007 June
with Megacam \citep{2006sda..conf..337M} at the 6.5\,m MMT.
The total integration time was 40 minutes under 
conditions of subarcsecond seeing.  The individual
exposures were reduced using standard techniques to 
create a final flattened and flux-calibrated mosaic covering
3C\,270.1.

\section{Data Analysis}
\subsection{Radio Morphology and Analysis}

\begin{figure}[ht]
\centering
\epsscale{0.8}
\hskip -3.0in
\includegraphics[height=2.5in]{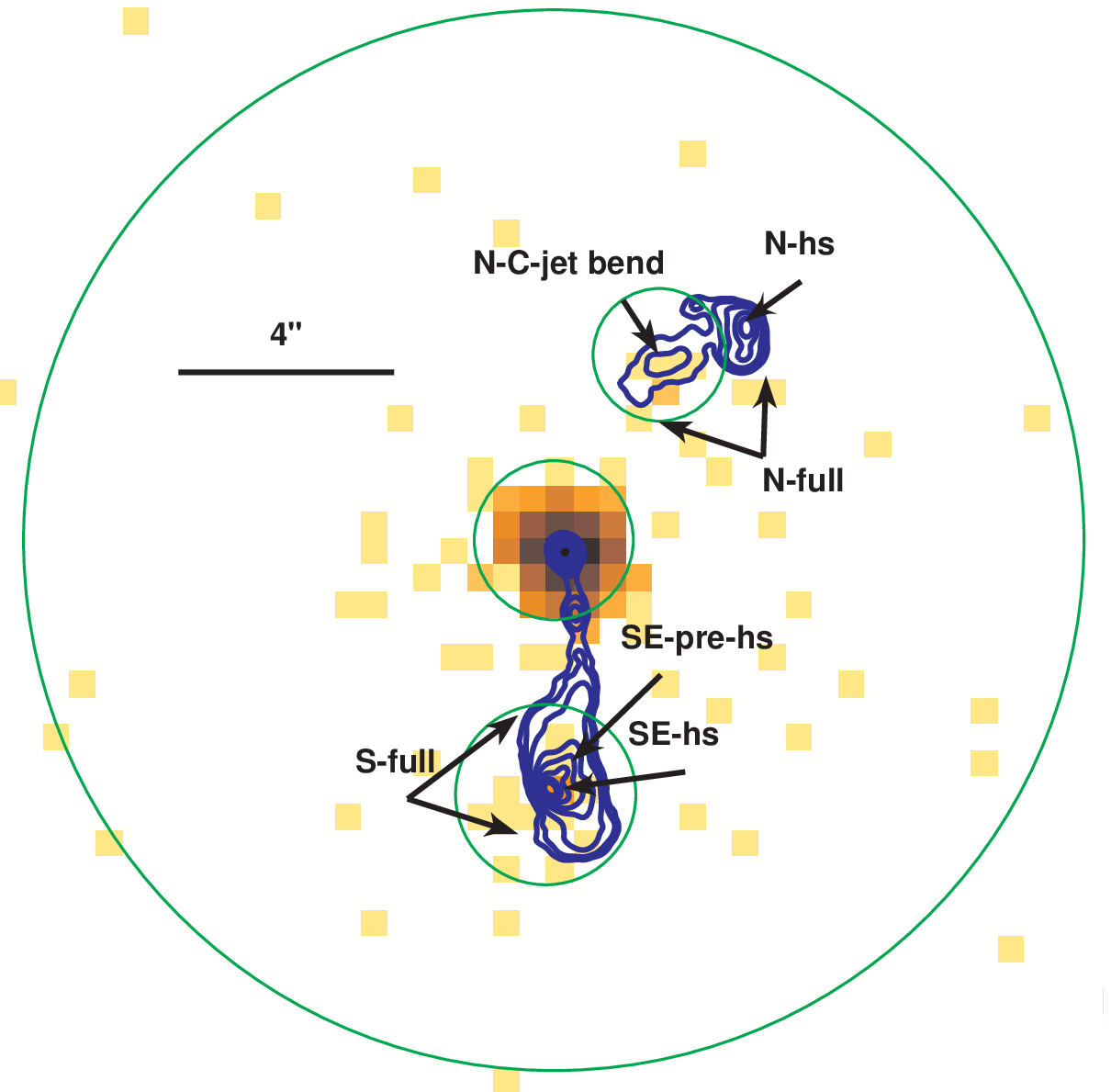}
\vskip -2.5in
\hskip 3.0in
\includegraphics[height=2.5in]{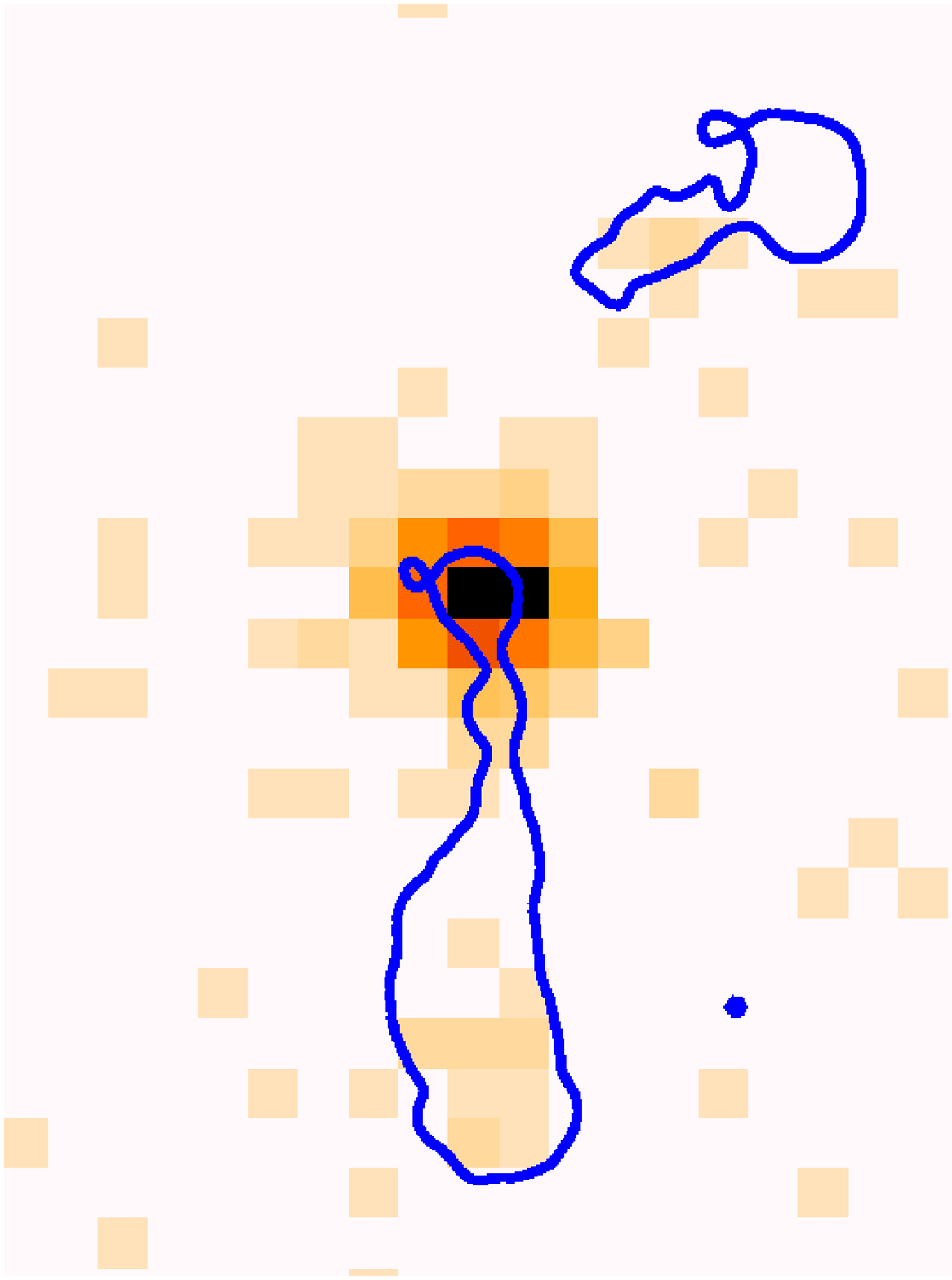}
\vskip -0.0in
\caption{\it 
Left: Radio contours in blue derived from the VLA radio map at 8.46 GHz 
(X-band) superposed on the unsmoothed, broad-band 
(0.3$-$8 keV) \chandra\ X-ray data at native (0.492~\arcsec) pixelation.
North is up and East to the left.  
Radio contour levels are: 
0.2,0 .36, 0.96, 5.12, 12.8, 25.6, 38.4,
54.4, 76.8 mJy beam$^{-1}$. 
The scale bar indicates 4\arcsec\
(34~kpc) and the outer circle has radius 10\arcsec. 
Key morphological features discussed in the text are
labelled.
The X-ray extraction regions for the quasar, the radio-related
emission north and south of 3C\,270.1 and the inner edge of the
background region annulus are shown in green.
Right: Zoom in on the central regions including 
the lowest radio contour level to highlight the 
radio-X-ray alignment.}
\label{fg:radio_X}
\end{figure}

An 8.46 GHz image of 3C\,270.1 
(contours displayed in Figure~\ref{fg:radio_X})
shows strong, unresolved emission from the central
quasar and more diffuse emission from the
lobes north and south of the nucleus with
some extension back towards the quasar nucleus.
The peaks of emission in the northern hotspot, the core, and the
southern hotspot are not co-linear.

\begin{figure}[ht]
\centering
\epsscale{0.6}
\plotone{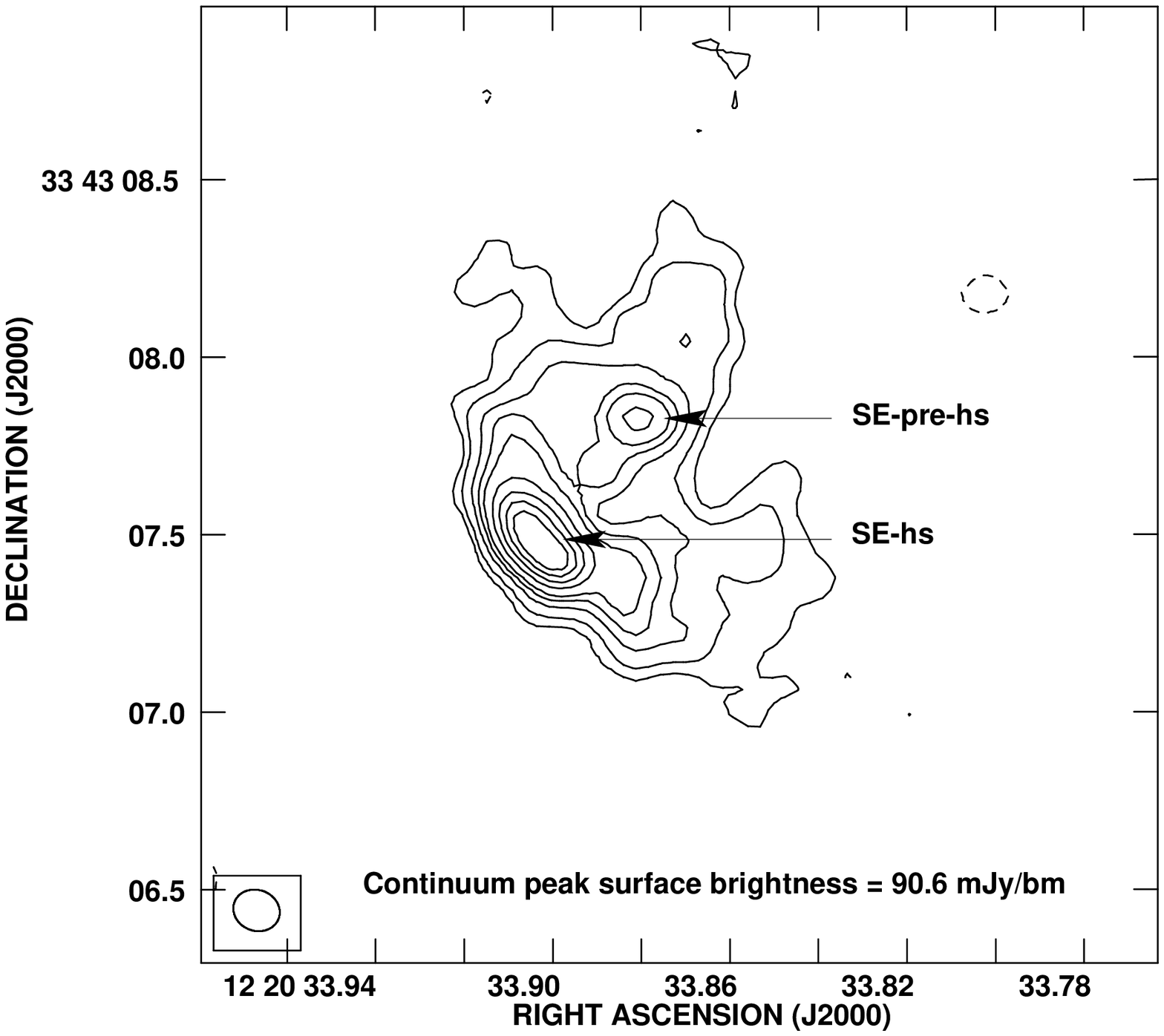}
\vskip -0.1in
\caption{\it 
Radio contour plot of the 14.9 GHz (U-band) VLA data for the southern
hotspot showing the morphology of the southeast (SE) and precursor
hotspots as described in the text. Contour levels are: $3 \times\ 0.22
\times\ (-1,1,2,4,6,8,12,16,20,24,32)$ mJy bm$^{-1}$, and
the beam is shown in the lower left corner.
}
\label{fg:radio}
\end{figure}

The southern emission includes a jet with position angle (PA) close to
180\deg\ 
and an extensive lobe at the same position angle extending over
$\sim 4\arcsec$ ($\sim 34$ kpc) which includes a hotspot at the
south-east edge. 
The jet includes a bright knot just south of the quasar
core. There is no apparent excess of X-ray counts at this position,
although the resolution of the X-ray data is sufficiently low
that a deconvolution would be challenging.
Figure~\ref{fg:radio} shows
higher-resolution 14.9 GHz radio contours of the southern hotspot
region. There is a double radio hotspot: 
 ``precursor'' and ``southeast (SE)'' components, with a pronounced
zig-zag morphology similar to 3C\,205 \citep{1984A&A...135...45L}.
A ridge of emission starts at the location of the precursor and
extends towards the peak of the southeast (SE) hotspot at a position angle 
of $\sim$ 140\deg. The ridge is brighter close
to the precursor hotspot in the 15~GHz image (Figure~\ref{fg:radio}) 
and becomes fainter farther away from it. 
From the peak of the SE hotspot, the ridge line turns sharply west, 
curving through $\sim$90\deg\ to a secondary peak and decreasing in 
surface brightness.  
The length of this curving ridge is 
about 1~\arcsec ($\sim$9~kpc).

The northern lobe is skewed by 30\deg\ from the axis defined by the
southern jet and consists of two distinct 
components \citep{1993ApJS...87...63L}. The 
well defined lobe emission includes a hotspot at the end farthest from the
quasar. The hotspot lies in an emission ridge with north-south
orientation and is approximately 0.\arcsec3 (2.6~kpc) in length. 
A faint radio emission component extends southeast from the lobe with
position angle $115^\circ$ and
is a relatively diffuse, steep-spectrum 
feature with length $\sim$ 2.\arcsec2 ($\sim$19~kpc). The feature turns
through an angle of $\sim$40\deg\
back towards the quasar, suggesting an encounter between an outflow of
radio plasma and a dense medium. We shall call this region the
counter-jet-bend (N-C-jet-bend). There is no jet emission visible on
the north side, and the lobe emission is significantly more 
depolarized \citep{1991MNRAS.250..171G}, 
further supporting the presence of hot gas surrounding the quasar
and consistent with this being the counter-jet side.

We used VLA X-band (8.46 GHz) and U-band (14.9 GHz) 
radio data (Figures~\ref{fg:radio_X}, \ref{fg:radio})
to estimate radio emitting volumes, flux densities, 
and spectral slopes in 
six regions defined based on the radio morphology:
N-hs (North hot spot), SE-hs (SouthEast hot spot), SE-pre-hs (SouthEast
precursor hot spot), 
N-C-jet-bend (North counter-jet-bend ), N-full 
(North lobe), and S-full (South lobe) (Table~\ref{tb:regions}).
These regions are also used as the basis for determining the optical, IR and
X-ray fluxes/limits, although the spatial resolution
is lower in the latter two 
bands so that hotspot- or lobe-related emission cannot be
separated.

\begin{table}[h]
\begin{center}
\caption{Summary of radio extraction
regions and flux densities and {\it Spitzer} upper limits
used in the SSC and iC/CMB modeling .}
\label{tb:regions}
\begin{tabular}{lccccrrrc}\hline
Region       & RA          & Dec         & Dimensions$^1$  & Angle$^2$
             & S$_{\small \rm 1.43}$ & S$_{\small \rm 8.46}$ &
             S$_{\small \rm 14.93}$ & Spitzer   \\
             & J2000       &  J2000      & of extraction
&            & mJy     & mJy    &  mJy   & Upper
             limits$^3$ $\mu$Jy\\
&&&region&&&&&3.6,4.5,5.8,8.0 $\mu$m \\ 
\hline\hline                                                                        
N-hs         & 12:20:33.61 & $+$33:43:16.1 & 0.16$''$$\times$0.09$''$
             & 00$^\circ$ &      & & 20 & \\
N-C-jet-bend & 12:20:33.73 & $+$33:43:15.1 & 1.0$''$$\times$0.7$''$
             & 30$^\circ$ & 60 &  7 &     & 2.1,3.6,23,27 \\ 
N-full       & 12:20:33.65 & $+$33:43:15.6 & 1.05$''$$\times$1.05$''$
             & 00$^\circ$ &570 & 65 &     & \\
SE-hs        & 12:20:33.90 & $+$33:43:07.5 &
             0.27$''$$\times$0.10$'' $
             & 120$^\circ$ &      &135 & 85  & 2.1,3.6,23,27 \\
SE-pre-hs    & 12:20:33.88 & $+$33:43:07.9 &
             0.22$''$$\times$0.16$''$  & 00$^\circ$ &      &  &
             10 & \\
S-full$^4$   & 12:20:33.87 & $+$33:43:07.8 & 2.1$''$$\times$0.8$''$
             & 00$^\circ$ & 1630 & 230 &     & 7.8,11,49,72 \\ \hline
\end{tabular}
\vskip 0.1in
\begin{minipage}{6.5in}
1: Semi-major $\times$ semi-minor axes of an ellipse, with volume
determined as a uniformly filled, prolate ellipsoid, except for S-full
where the dimensions are the height and radius of a right circular cylinder. \\
2: The angle of rotation of the major axis to the east of north. 
 \\
3: 3.6$\sigma$ upper limits for a point/extended source as appropriate \\
4: Radio flux for the full lobe with the hotspot emission subtracted \\
\end{minipage}
\end{center}
\end{table}

\subsection{X-ray Data Analysis}

The unsmoothed, broad-band, X-ray data are shown in 
Figure~\ref{fg:radio_X} with the radio contours superposed.
The X-ray spatial resolution is $\sim 0.\arcsec5$, 
about twice that of the radio data.
The X-ray data were shifted 0.\arcsec32 to the SW (\chandra's 
offset has 0.\arcsec4 rms
radius, Proposers' Observatory Guide) 
to align the peak of the nuclear emission with the VLA position of the
radio core. 

The analysis of the data from several regions in the vicinity of
3C\,270.1 is described below along with
that of a lower-redshift ($z$=1.038) quasar NGC 4395 B06, 
$\sim 48 \arcsec$ north
of 3C\,270.1 (on the same ACIS-S chip and with similar point spread function), 
which is used as a comparison point source.
Extracted counts and derived fluxes 
are listed in Table~\ref{tb:counts}.

\begin{figure}[ht]
\centering
\epsscale{1.1}
\plottwo{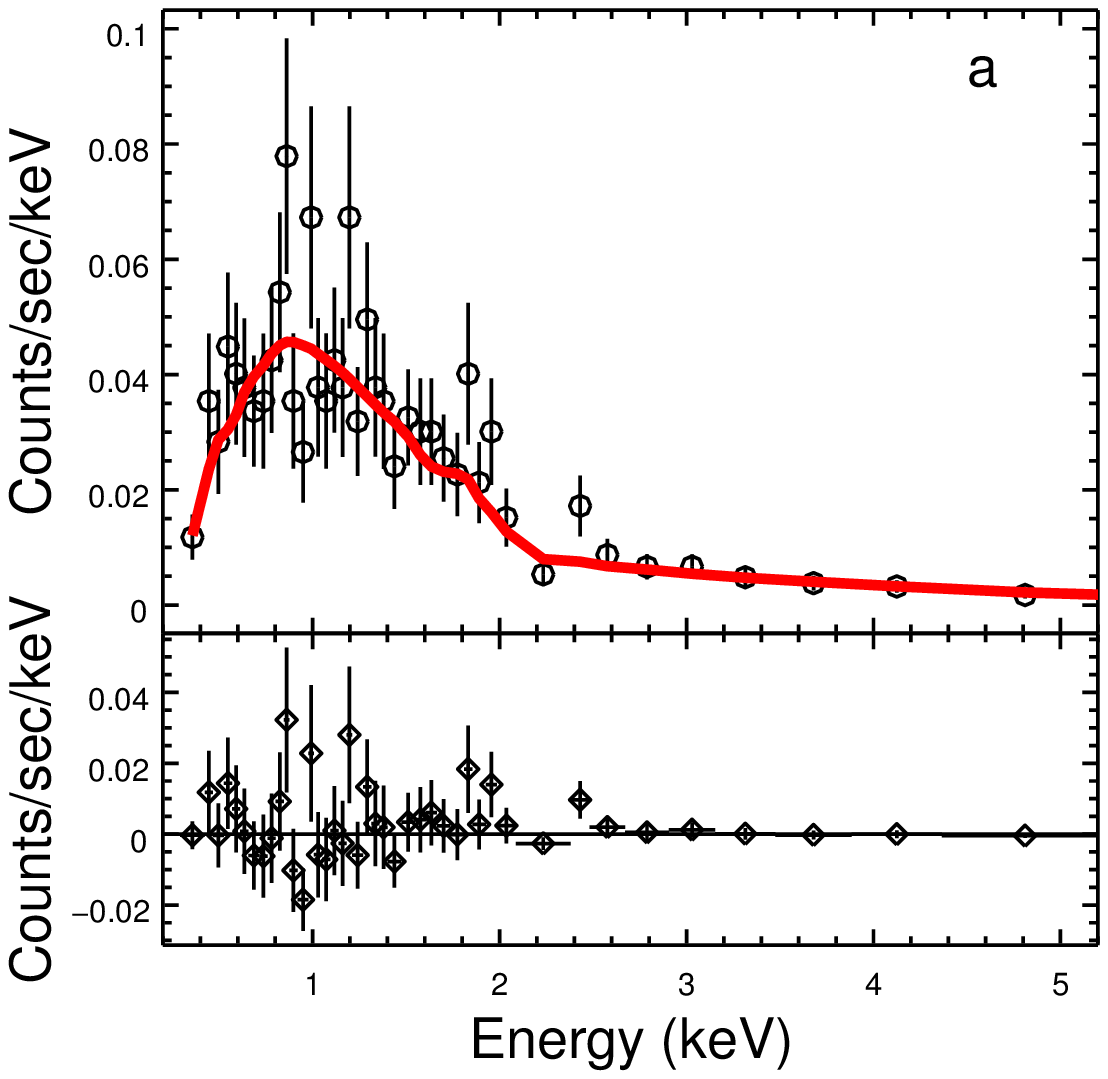}
{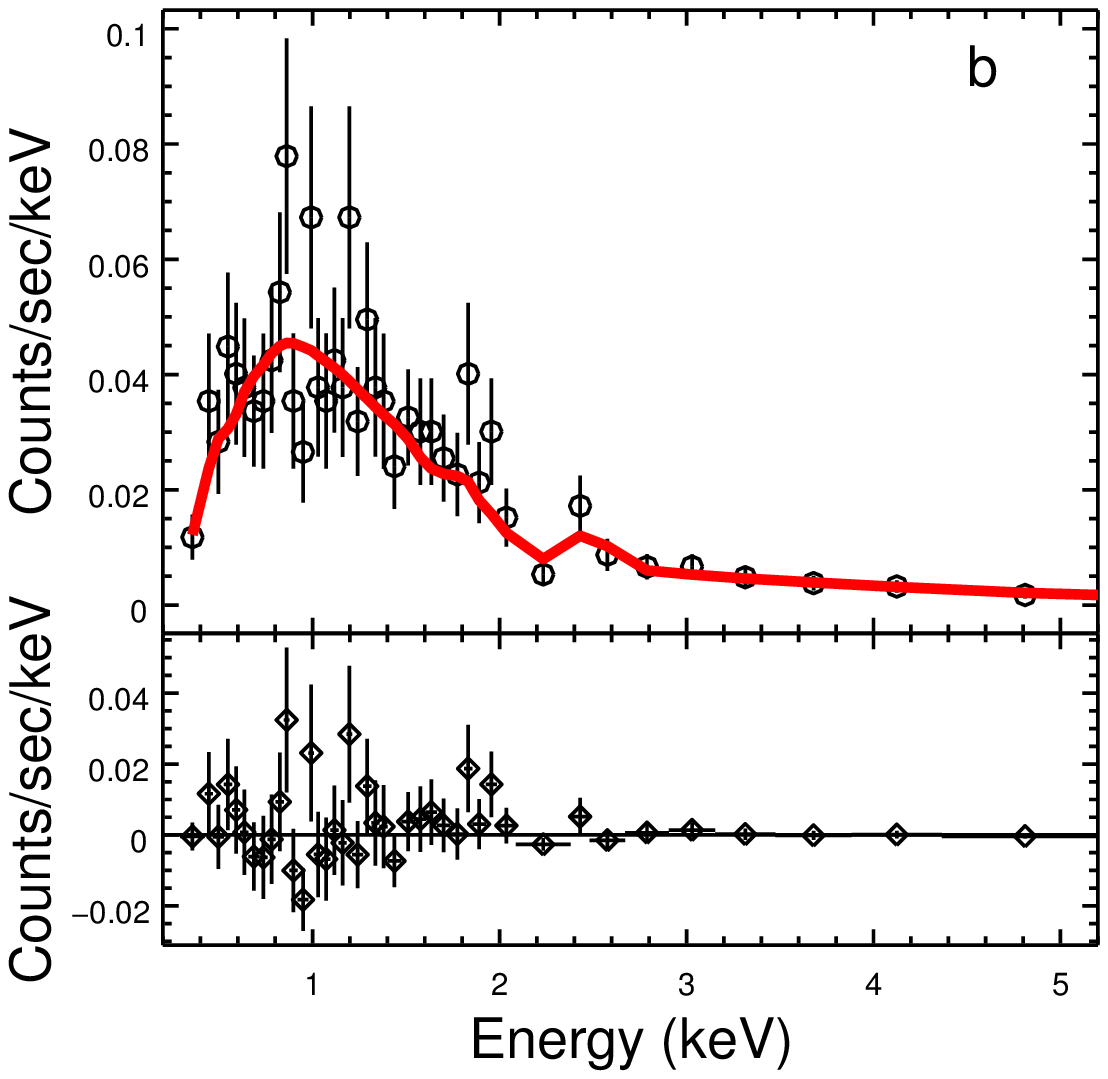}
\vskip -0.2in
\caption{\it a: \chandra\ X-ray spectrum of 3C\,270.1 with the best fit
  power-law model ($\Gamma=1.64\pm0.08$, Galactic \nh) superposed
  and the residuals below. Deviations in the residuals at
  $\sim 2.5$ keV suggest the
  presence of an emission line are visible. b: As for a, but 
including an Fe K$\alpha$ emission line (EW=172$\pm$110 eV) fixed at energy  
$2.5$ keV, consistent with Fe K$\alpha$ from cold
  material at the redshift of the quasar. 
}
\label{fg:AGNfit}
\end{figure}

\begin{center}
\begin{deluxetable}{lcccccc}
\small
\tablecaption{X-ray counts and derived X-ray parameters$^1$ 
\label{tb:counts}}
\tablehead{
  \colhead{Region Name} & \colhead{Net Counts}  &
  \colhead{HR$^2$} & \colhead{Flux Density$^3$} & \colhead{Flux$^4$} &
  \colhead{Flux$^4$} & \colhead{Flux$^4$} \\
\colhead{} & \colhead{0.3$-$8keV} &\colhead{} &\colhead{1keV}&
  \colhead{0.3$-$8keV} &\colhead{0.3$-$2keV} & \colhead{2$-$8keV}
}
\startdata
Quasar Core & 726 $\pm$ 27 & $-0.53\pm$0.03 & 1.4 & 6.0 & 2.5 &
  3.5  \\
Southern Hotspot$^5$ & 16.7 $\pm$ 4.1 & $-0.86^{+0.04}_{-0.14}$ &
0.034 & 0.111 & 0.062 & 0.049  \\
N-C-jet-bend$^5$ & 5.8 $\pm$ 2.5 & $-0.9\pm0.1$  & 0.008 &
0.027 & 0.015 & 0.012 \\
Diffuse Emission$^6$ &  22.8$\pm$5.6 & $-0.09\pm$0.22 &
0.05 & 0.13 &  0.09 & 0.04  \\ 
NGC 4395 B06 & 351 $\pm$ 19 & $-0.63\pm$0.04 &  0.8 & 2.5 & 1.4
& 1.1 \\
\enddata \\
\vskip 0.1in
\begin{minipage}{6.5in}
1: Fluxes are determined using the best fit model in
Table~\ref{tb:fit} unless otherwise noted. Errors are quoted at 
the 1$\sigma$ level. \\
2: Hardness ratio, HR=(H-S)/(H+S) (H(2$-$8 keV), S(0.3$-$2 keV)), 
determined using the BEHR method 
\citep{2006ApJ...652..610P} \\
 3: in units of $10^{-13}$ erg cm$^{-2}$ s$^{-1}$ keV$^{-1}$\\
 4: in units of $10^{-13}$ erg cm$^{-2}$ s$^{-1}$, statistical errors
 in counts dominate for the extended emission \\
 5: Assuming $\Gamma=2$ and Galactic \nh=1.29 $\times 10^{20}$
 cm$^{-2}$. Southern hotspot includes both SE-hs and pre-hs 
which are not resolved in the X-ray data \\
 6: APEC model (Astrophysical Plasma Emission
 Code)\footnote{http://www.atomdb.org/} 
for collisionally-ionized thermal plasma
assuming kT=4 keV; fluxes and flux density are 
corrected for the 28\% excluded area encompassing the radio-linked
X-ray emission.
\end{minipage}
\end{deluxetable}
\end{center}

\subsubsection{The Quasar Core}

The quasar core is unresolved and well-detected. 
Counts were extracted from a 1.5$''$ radius, 
circular region centered on the quasar position
($\alpha$\ = 12\,20\,33.9; 
$\delta$\ = +33\,43\,12), and background counts, estimated 
from an annulus $10''-20''$ with the same center, were subtracted yielding 
726$\pm$27 (0.3$-$8 keV) net counts (Table~\ref{tb:counts}). 
A point source correction was applied for flux estimates using the
CIAO tool arfcorr, which applies an energy-dependent PSF correction 
appropriate for the extraction region to the ARF (effective area) file.
The counts were grouped to yield a minimum of 15 per bin to perform
spectral fitting. 
A single, absorbed power-law spectral fit in the energy range
(0.3$-$8 keV) shows absorption consistent with the Galactic 
\nh~=1.29$\times 10^{20}$ cm$^{-2}$ \citep{1992ApJS...79...77S}, 
which was accordingly fixed to
this value. The resulting best fit power-law slope, 
$\Gamma = 1.64\pm0.08$ ($\chi^2_{red} \sim
0.74$, 42 degrees of freedom (dof)) is consistent with those generally
reported for high-redshift, radio-loud quasars
(e.g. \citet{2011ApJ...738...53S,2006MNRAS.366..339B},
\citet{2005ApJS..156...13M}, Figure~\ref{fg:AGNfit}a, 
\citet{1997ApJ...478..492C}).
There is an emission line apparent at energy, E$\sim 2.5$ keV, 
consistent with cold Fe K$\alpha$ 
in the rest frame of the quasar. Although only marginally significant, 
addition of a narrow Gaussian line 
at 2.5\,keV improves the fit, 
while a higher energy line (2.7\,keV, approximating ionized Fe) 
is clearly inconsistent.
A spectral fit including a narrow line with energy, E$= 2.5$ keV
yields an EW=$172 \pm 110$\,eV with a
power-law slope $\Gamma = 1.66 \pm 0.08$ ($\chi^2_{red} \sim
0.70$, 41 dof). Given the large errors, 
the line energy and equivalent width are 
consistent with those reported for detected Fe K$\alpha$ emission in 
radio-loud quasars (\lax 100 eV, \citet{2006ApJ...642..113G}).
There is a suggestion (Figure~\ref{fg:AGNfit}b)
of further emission lines around 0.8 and 0.5 keV
which roughly align with highly-ionized Si and Mg features in a
photoionized spectrum in the quasar rest-frame. However they are not 
significant in this low-count spectrum. 
The fit parameters with and without the Fe K$\alpha$ 
emission line are given in Table~\ref{tb:fit}.
Consistency of the data with the Galactic \nh\ indicates no evidence for
intrinsic absorption or the presence of a soft excess component, nor is there
any improvement to the fit if a reflection component is added. 
Both intrinsic absorption and reflection components tend to be weak
in face-on, type I quasars, particularly those that are radio-loud
such as 3C\,270.1 \citep{2008MNRAS.390.1217M, 2011ApJ...738...53S}

\begin{deluxetable}{lllcccc}
\centering
\tablecaption{Parameters for X-ray (0.3$-$8 keV) Spectral Fits$^1$ 
\label{tb:fit}}
\tablehead{
  \colhead{Region Name} & \colhead{Model} &
  \colhead{$\Gamma$} & \colhead{Line E} &
  \colhead{Line EW$^2$ } & \colhead{$\chi^2$} & \colhead{dof$^3$}  \\
\colhead{} & \colhead{}  & 
& \colhead{keV} & \colhead{eV} & \colhead{} & \colhead{} }
\startdata
Quasar Core$^4$ & PL+Gaus &  1.66$\pm$0.08 &2.5& 
172$\pm$110 & 28.7 & 41\\
& PL& 1.68$\pm$0.08 &$-$& $-$ & 31.3 & 42 \\
NGC 4395 B06 & PL & 1.99$\pm$0.15 & $-$ & $-$ & 29.4
& 21 \\
\enddata \\
\vskip 0.1in
\begin{minipage}{6.5in}
1: Assuming Galactic \nh\ = 1.29$\times 10^{-20}$ cm$^{-2}$. Errors
are quoted at the 1$\sigma$ level for one interesting parameter
(unless otherwise noted)
 \\
2: Equivalent Width of emission line \\
3: Data were re-binned with a minimum of 15 counts per bin\\
4: Errors are 1$\sigma$ for 2 interesting parameters.
\end{minipage}
\end{deluxetable}

\subsubsection{Radio--lobe-associated X-ray emission}
\label{sec:dlal}
X-ray emission is present coincident with the southern
radio lobe, and the peak is close to the primary (SE) radio hot spot.  
The net counts were extracted from a circle of radius 1.7''
centered on $\alpha$ = 12\,20\,33.91, $\delta$ = +33\,43\,07.25, which
includes the cluster of counts in this region and aligns well with the
position of the double radio hotspot. 
The net counts, 16.7$\pm$4.1, are primarily soft, with hardness ratio
HR=$-0.86^{+0.04}_{-0.14}$. There are too
few counts for a spectral fit. Fluxes (Table~\ref{tb:counts})
are insensitive to the assumed spectral form since statistical
uncertainties dominate.

X-ray emission north of the quasar is centered south-east of the radio
hotspot in the 
northern radio lobe, aligned with the position of the N-C-jet-bend in the
radio emission (Figure~\ref{fg:radio_X}). 
Counts were extracted from a
circular region, radius 1.\arcsec25, 
centered on $\alpha$= 12:20:33.74, $\delta$
= 33:43:15.4, selecting the cluster of events apparently associated with
the radio emission in this region of sky. We find 6 counts where
0.2 background counts are expected, a significant detection.
The net counts, 5.8$\pm$2.5, in the broad-band (0.3$-$8
keV) are soft with all the counts below 2 keV giving a hardness ratio, 
HR=$-0.9\pm0.1$ (Table~\ref{tb:counts}). 

There is no detected X-ray emission associated with the northern radio hotspot.
The more extended lobe emission possibly includes faint X-ray
emission at its southern edge, but the association is 
highly uncertain given the low signal-to-noise and the lack of 
alignment with the lobe. 

\subsubsection{Nearby X-ray source: NGC 4395 B06.}

There is a strong point-like X-ray source 48\arcsec\ off-axis, close enough
that there is no significant change in the PSF, on the same
ACIS S-3 chip, and located at the
position of the radio-quiet quasar NGC 4395 B06 ($\alpha$ = 12 20
32.7; $\delta$ = +33 43 56, $z$=1.038, 
\citet{2001AJ....121.2843B}). Extracting counts
from a circular region of radius 1.5$''$, as for 3C\,270.1, and
estimating background counts from a $5''-10''$ annulus centered on the
quasar, yields
351$\pm$19 net counts. The hardness ratio, HR=$-0.63 \pm 0.04$ is
similar to that of the nucleus 
of 3C\,270.1 (Table~\ref{tb:counts}). A power-law fit assuming Galactic
\nh\ yields $\Gamma = 1.99 \pm 0.15$ ($\chi^2_{red}=0.89$, 21 dof, 
Table~\ref{tb:fit}). Fitting with \nh\
free results in a slightly flatter power-law slope 
($\Gamma = 1.92 \pm 0.14$, $\chi^2_{red}=0.87$, 20 dof) and very low \nh\
suggesting the presence of a soft excess. However the signal-to-noise is
insufficient to distinguish between models for a soft excess and the 
power-law slope is typical of radio-quiet quasars \citep{2005A&A...432...15P}
so no more complex modeling was carried out.
Fluxes were determined using the initial fit, assuming Galactic \nh.


\begin{figure}[h]
\centering
\centering
\epsscale{0.6}
\plotone{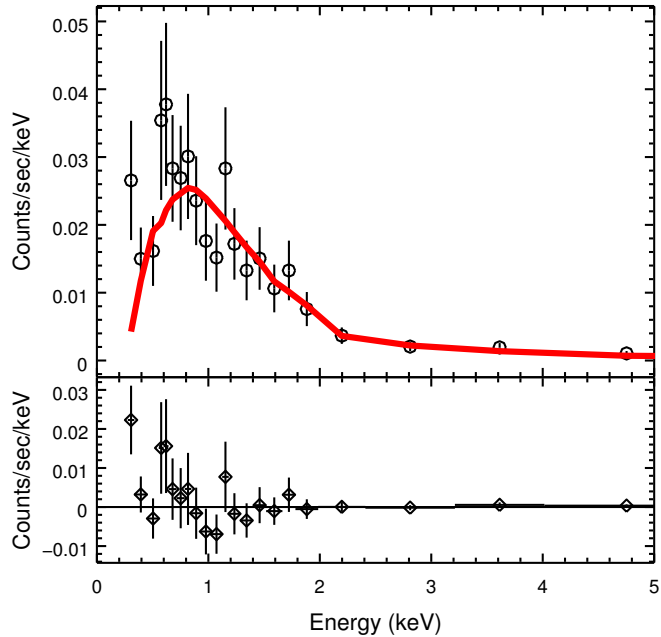}
\vskip -0.2in
\caption{\it The best-fit single power-law ($\Gamma=1.99\pm0.15$,
 Galactic \nh )  
 spectral model for NGC4395 B06 superposed on the X-ray data
  with the residuals
  below.}
\label{fg:NGC4395bf}
\end{figure}

\subsubsection{Diffuse Extended Emission around 3C\,270.1}
\label{sec:cluster}
There are additional X-ray counts around the quasar core
which are not associated with the envelopes of radio emission.
The emission occurs on both sides of the quasar but the number of
counts is too low to constrain its spatial distribution.

We extracted counts in the broad band (0.3$-$8
keV) from an annular region centered on 3C\,270.1
and extending from $2\arcsec-7.\arcsec5$ ($\sim$64 kpc)  excluding  
two segments (opening angles: North: 60\deg , 
South: 40\deg) containing the radio-related regions, 
this excluded 28\% of the annulus. 
Background counts (34) were estimated from an annular
region extending from $10''-20''$. 
Twenty-seven counts were detected where
4.3$\pm$0.7 are expected, a detection which is highly significant
(Table~\ref{tb:counts}).
The net extended counts are 22.8$\pm 5.6$ (0.3$-$8 keV) with 
an energy distribution harder 
(HR = $-0.09\pm0.22$) than the quasar or the radio-linked X-rays.
Extracting counts from a similar extended 
region ($2.\arcsec5-7.$\arcsec5) 
around the bright, nearby quasar NGC4395 B06 yields only 9 counts where 
5$\pm1$ are expected, a marginal detection of 3.9 counts. 
We thus conclude that $< 20$\% of
the diffuse extended counts around 3C\,270.1 
could originate in the wings of the central quasar.
Since the wings are more extended at soft energies, 
this would result in an even harder spectrum for the diffuse emission. 

The hardness of the diffuse counts makes it
unlikely that they are nuclear X-rays Thomson-scattered 
from an extended gaseous
halo. They could be CMB photons Compton-scattered by electrons
associated with fainter, more
diffuse radio emission below the current radio flux limit, as
suggested for the
high-redshift radio source 3C 294 \citep{2003MNRAS.341..729F}, or 
thermal emission from hot gas in the optical/IR cluster
\citep{2009ApJ...695..724H}. 

Since the mean surface brightness of the diffuse emission 
is 0.18 cts arcsec$^{-2}$, it might contribute 1.6 and 0.9
counts in the extraction regions for the southern
hotspot and northen N-C-jet-bend, 6\% and 15\% of the detected counts
respectively, assuming it is isotropic. 
Correction of the X-ray spectra of the radio features 
would result in even softer emission and
somewhat lower fluxes in these regions (Table~\ref{tb:counts}).
Any contribution to the quasar core would be negligible ($\sim 1.3$ of the 
726 net counts).

\subsection{{\it Spitzer} and Optical, $z^\prime$ Data Analysis}

\begin{figure}[ht]
\centering
\epsscale{1.0}
\plottwo{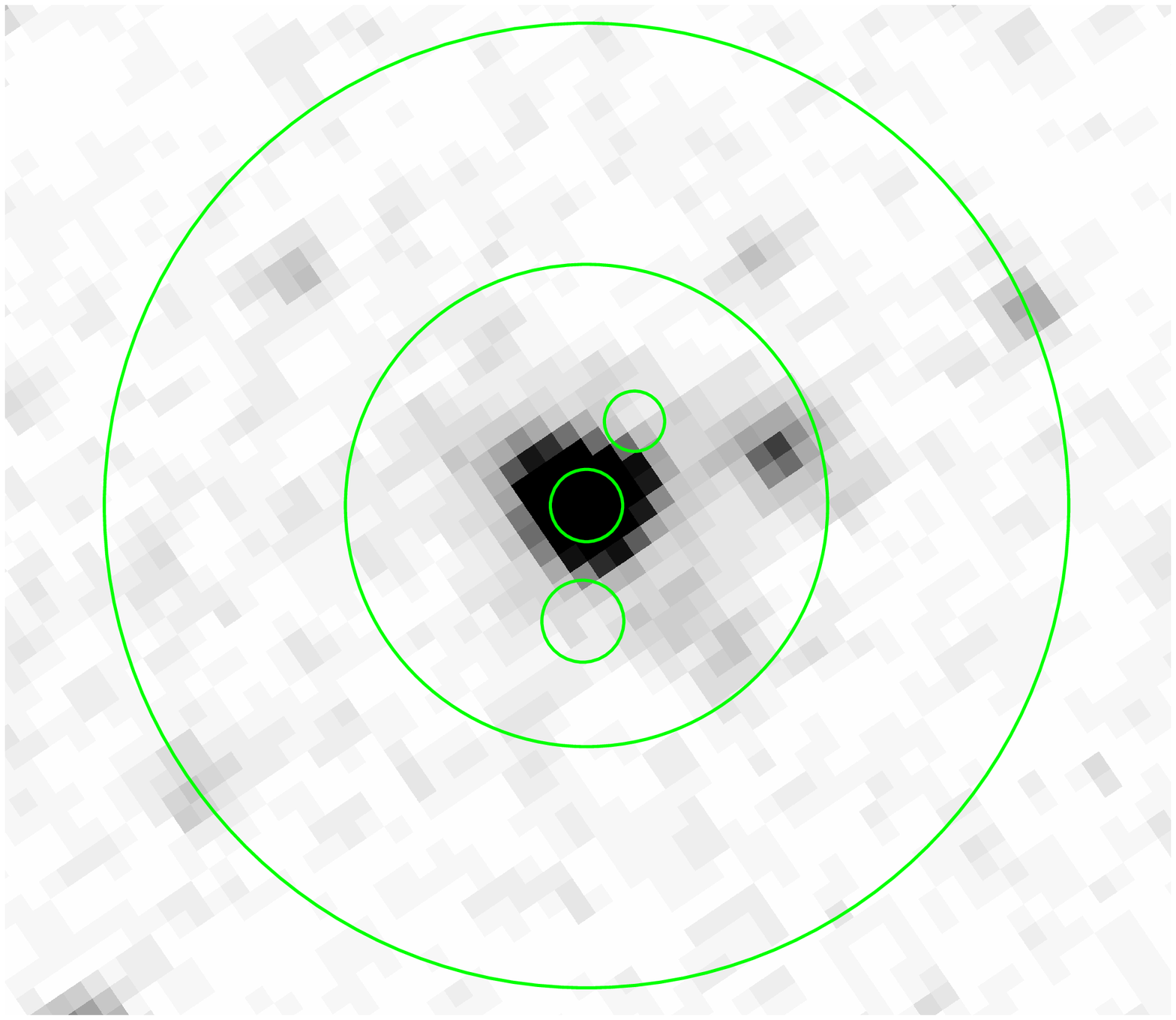}{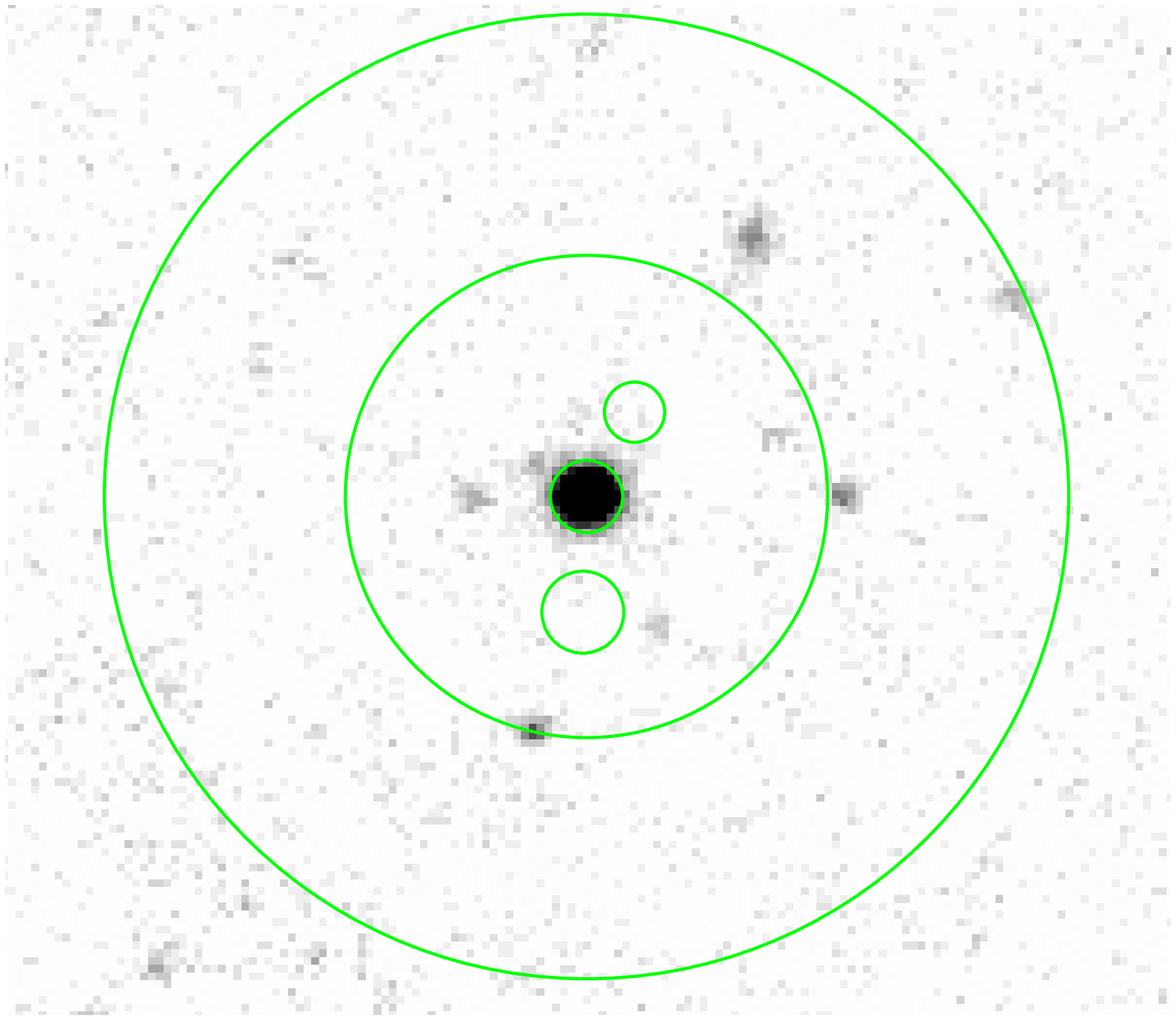}
\caption{\it Negative images in the IRAC 4.5 $\mu$m (left) and optical
 $z^\prime$ bands of 
  the sky region around 3C\,270.1 with north up and east to the
  left. The X-ray extraction regions for 
  the quasar core, the northern and southern radio structures, and the
  background  are 
  superposed. The outer circle has radius 20$\arcsec$. 
  There is no detectable IR or optical emission associated with the 
  radio lobes. }
\label{fg:IRAC}
\end{figure}

The 4.5 $\mu$m IRAC data for 3C\,270.1 are shown in
Figure~\ref{fg:IRAC} with the X-ray extraction regions superposed. 
The quasar is well-detected in the IRAC data. 
There is no detected emission associated with either of the radio lobes.
Upper limits were determined for the radio
hotspots and the N-C-jet-bend appropriate for a point source 
since any associated IR emission would be unresolved with {\it Spitzer}.
A second set of upper limits, appropriate for extended IR emission
features, was determined for the full southern radio lobe based upon
the radio contours.
The upper limits are given in Table~\ref{tb:regions}
and used to constrain the spectral energy distribution models
described in Section~\ref{sec:model}.

The $z^\prime$ optical image is also shown in Figure~\ref{fg:IRAC}
with the X-ray extraction regions superposed. There is no detectable optical
emission at the position of the radio lobes. Upper limits were
determined appropriate to  the X-ray regions. 
We estimated the sensitivity 
from the fluctuations measured in source-free pixels 
to be 21.8 AB mag (6.9~$\mu$Jy, $5\sigma$) within a 3\arcsec\ diameter
aperture.

There is a faint source $\sim 6$\arcsec\ east of 3C\,270.1, also
marginally visible in the IRAC data. It is most likely a faint
galaxy. The position is $\sim 1.$\arcsec4 east of a small group of
X-ray photons currently identified as part of the extended, diffuse
emission. Given this offset, the X-ray counts are unlikely to be related, but
deeper \chandra\ X-ray data would clarify any possible association.

\section{Modeling the X-Ray Emission from Hot Spots and Radio Lobes}
\label{sec:model}

There are three mechanisms commonly invoked to explain 
radio-associated, extended X-ray emission: synchrotron, SSC and
iC/CMB. All three were applied to the components of 3C\,270.1 following the
method described by \citet{1998MNRAS.294..615H}. 
Given the lower spatial resolution and low signal-to-noise of the
\chandra\ data, we are unable to uniquely identify the X-ray emission
with specific radio structure based on spatial distribution alone.
We therefore determined the fluxes from the north and south
radio structures in regions appropriate for the observed X-ray emission
and constructed radio through X-ray spectral energy distributions 
for the radio regions assuming, in each case, that all
the X-ray emission is associated with the particular radio region
being considered. The models were then applied to each SED.

The radio through X-ray SED for each region was fitted with a standard
continuous injection electron spectrum with a minimum energy corresponding
to $5 \times 10^6$ eV and maximum energy corresponding to 
$5 \times 10^{12}$ eV for the S-full and
N-C-jet-bend.  For the SE-hs, we used the identical minimum
energy but a maximum energy of $5 \times 10^{10}$ eV.  
The choice of high energy cutoff was due to the following two
constraints: (i) it is as low as possible without being inconsistent with
the radio data and (ii) the fitted synchrotron model to these data does not
violate the {\it Spitzer} upper limits.  
There remain significant uncertainties in the maximum electron energy,
but for $\alpha > 0.5$ and E$_{max} >>$ E$_{min}$ the derived physical
parameters are not
very sensitive to this value. 
We assume that non-radiating particles are
insignificant in the radio plasma and use a  filling factor of 1.

The predicted X-ray flux density for each region was determined for
respectively, 
a broken power-law synchrotron model with a high-energy cutoff, a synchrotron
self-Compton model (SSC) in which the X-ray emission is the result of
iC scattering of the synchrotron radio photons off the relativistic
electrons, and a model of iC
scattering of CMB photons (iC/CMB)
off the radio synchrotron-emitting, relativistic electrons. 
The models and predictions were determined in the quasar's rest
frame and are shown in
Figures~\ref{fg:south} and \ref{fg:northfaint} superposed on the
observed SED. 
The infrared upper limits (at 3.6$\sigma$) for each region from the Spitzer
IRAC data sometimes constrain the highest energy electron populations.
The figures also include a $z^\prime$ upper limit
appropriate for a point source, but this does not
provide an independent constraint.

\subsection{Southern Radio Lobe/Hotspot}

\begin{figure}[h]
\centering
\epsscale{1.0}
\plottwo{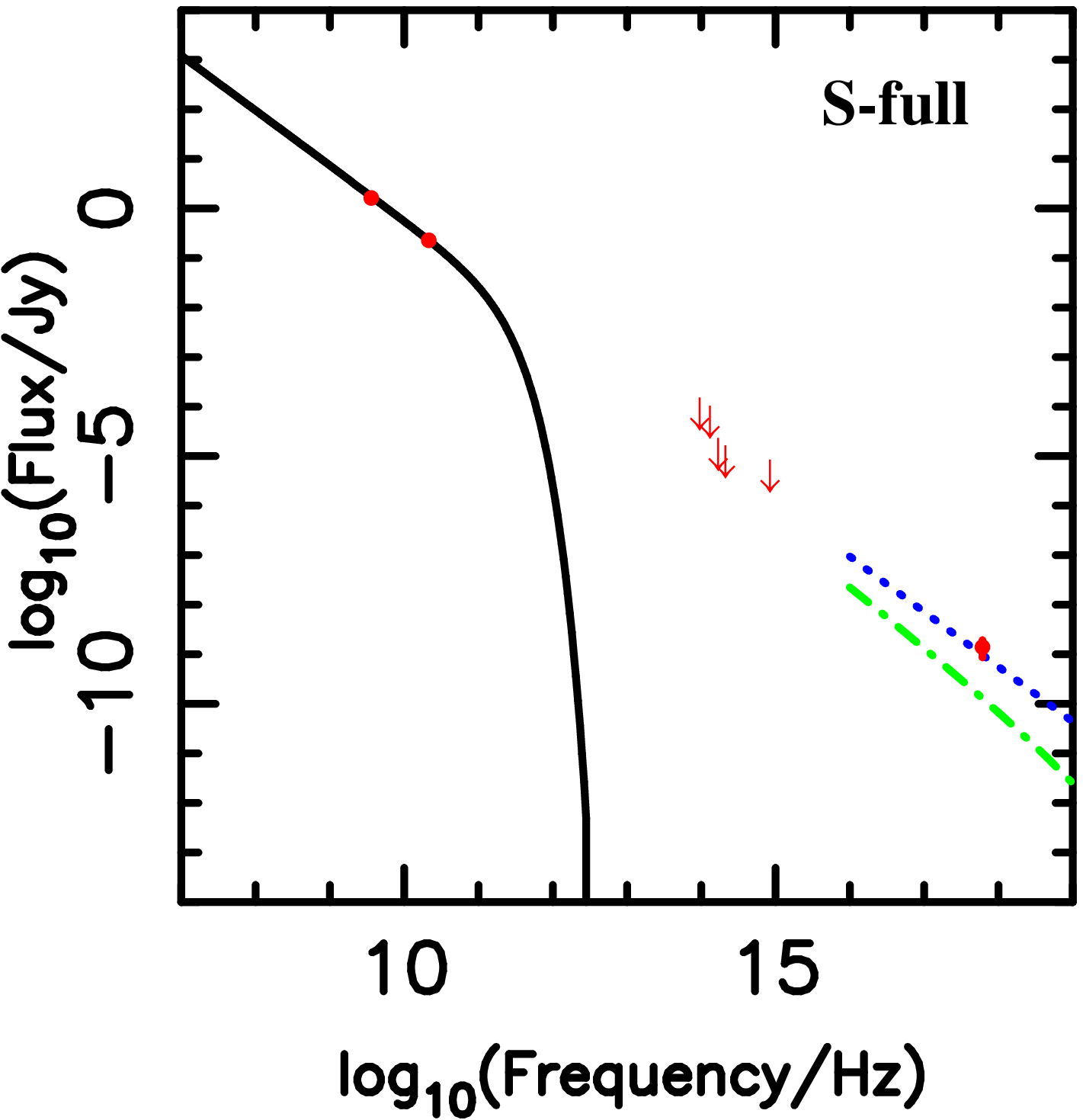}{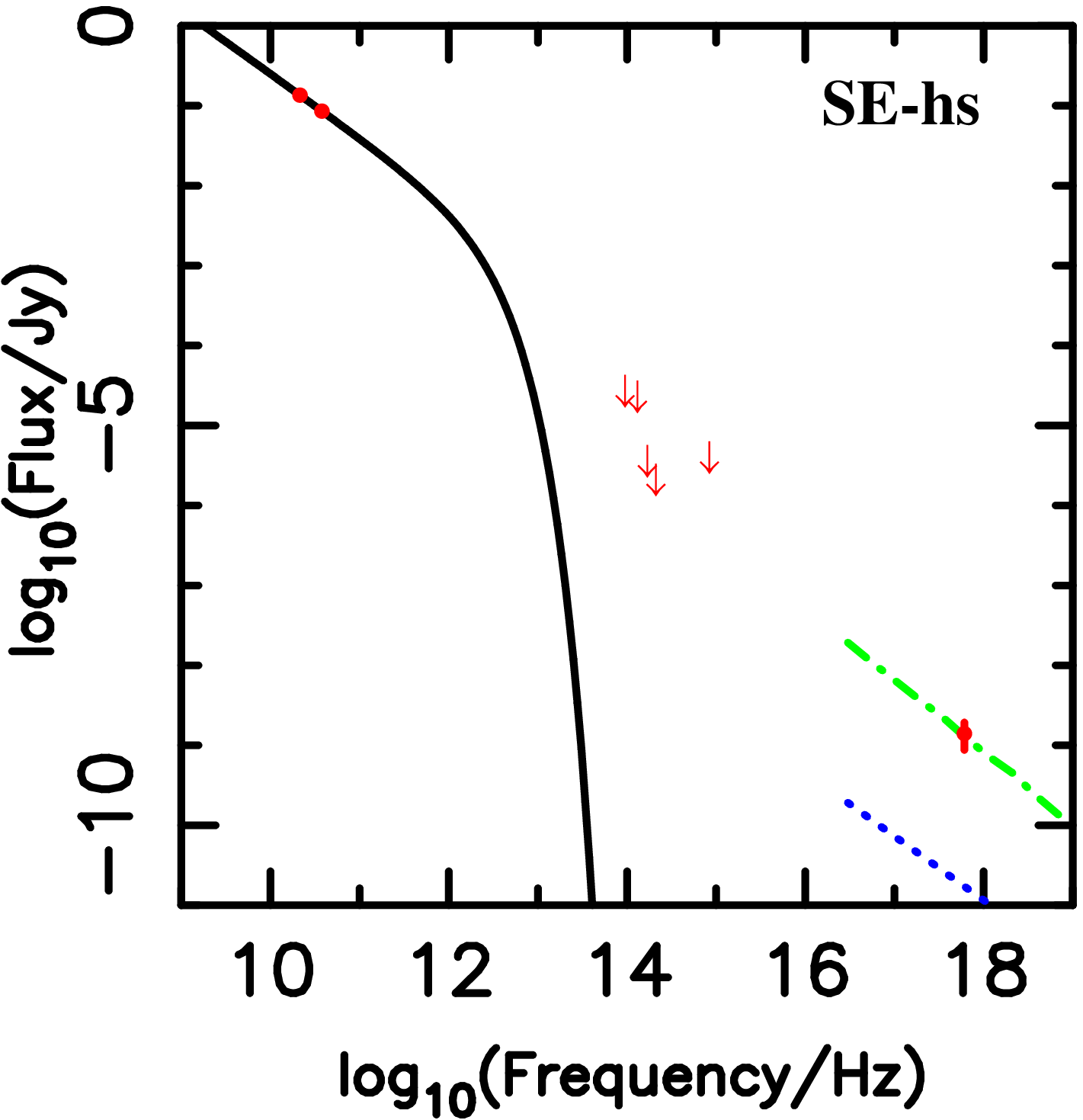}
\caption{\it Left: Spectral Energy Distribution (SED) for the Southern
  radio lobe (S-full, Table~\ref{tb:regions}).
The synchrotron model consistent with the radio data (black line)
and the predicted X-ray emission 
from SSC (green dash-dot) and iC/CMB (blue dotted) 
for a magnetic field of 8\,nT are superposed.
The model is based on the VLA data for the lobe
with the flux density of the hotspot subtracted.
Right: The SED for the south-east component of the double hotspot 
(SE-hs in Table~\ref{tb:regions}) with
the synchrotron model and predicted X-ray emission levels superposed.
SSC emission with a field of 24\,nT (roughly a third 
of the equipartition field) is
consistent with the data. In both figures the data and model are
blueshifted to the source frame, the 
red data points show the radio and X-ray flux densities and red
arrows show the {\it Spitzer} and $z^\prime$ upper limits. 
The error bars indicate 1~$\sigma$ statistical uncertainties, which 
are smaller than the radio data points. 
}
\label{fg:south}
\end{figure}

The signal-to-noise of the \chandra\ data is too low 
for us to determine whether
the X-ray emission arises from the entire southern 
radio lobe or the hotspot(s). 
If the X-ray emission originates throughout the lobe, the 
synchrotron and SSC predictions are 1-2
orders of magnitude below the observed X-ray flux while the iC/CMB
prediction is consistent with the observed X-ray flux if the magnetic
field is set at 8\,nT, about a quarter of 
the equipartition field for this region 
(Figure~\ref{fg:south}, left).

If instead all the X-ray emission
originates in the SE hotspot, as suggested by the alignment of the peak
emission, the {\it Spitzer} and optical upper limits rule out direct 
synchrotron as a viable mechanism for the emission at IRAC and shorter
wavelengths, including the X-ray emission. The iC/CMB prediction
for this small region is $\sim$3 orders of magnitude too low. The 
SSC prediction with a minimum-energy, equipartition
magnetic field of 86\,nT is also low but can be brought into agreement
with a field of 24\,nT (Figure~\ref{fg:south}, right), 
about a third of this value. This is not
an unusual departure \citep{2004ApJ...612..729H} and not surprising given 
the uncertainties involved. 
For example, the X-ray emission could originate in
more than one component and allow
for a model closer to the equipartition field.

The radio emission from the precursor hotspot region 
(SE-pre-hs, Table~\ref{tb:regions}) can only be determined at one
frequency. Given the lower spatial resolution of the {\it Chandra} 
data and the lack of constraint on the 
optically-thin part of the radio spectrum and subsequent extrapolation
of the electron spectrum to low energies, we cannot
rule out a significant contribution from SSC
emission from the precursor region.

In summary, 
the most likely interpretation, given the alignment between the X-ray
emission and the radio emission, is that the X-rays primarily originate
in the hotspot and/or the pre-cursor hotspot 
and are due to SSC. For emission from only the SE hotspot, 
the field is $\sim$ 24\,nT, about a third of the equipartition field. 

\subsection{Northern Radio Emission}

There is no detectable X-ray emission aligned with the northern hotspot.
The SSC X-ray prediction  for the northern hotspot,
for an equipartition field of 77\,nT based on the radio flux,
assuming $\alpha = 0.7$ and 
constrained by the IR upper limits, is about a factor of 10 lower 
than the X-ray upper 
limit (1.3$\times 10^{-4} \mu$Jy, $3 \sigma$) so that no X-ray
emission is expected. 

\begin{figure}[h]
\centering
\epsscale{0.5}
\plotone{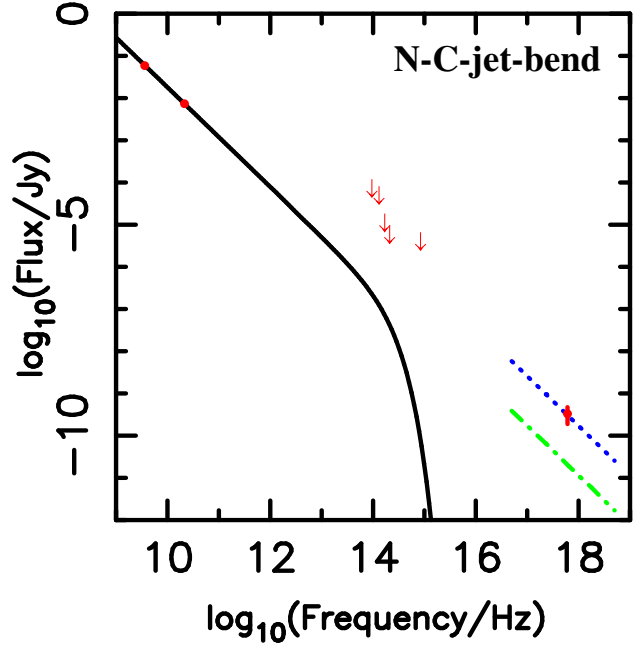} 
\caption{\it The SED for the N-C-jet-bend region, on the east side of the
  northern radio lobe (Table~\ref{tb:regions}) with the direct
  synchrotron component matched to the VLA data (black)
  and the predicted SSC (green) and iC/CMB (blue) X-ray
  emission adopting a magnetic field of 3$\pm1$\,nT, a factor of $7-10$
  lower than the equipartion field. The data
and model are blueshifted to the source frame, symbols are as in 
Figure~\ref{fg:south}.}
\label{fg:northfaint}
\end{figure}

There is significant X-ray emission aligned with the N-C-jet-bend
(Table~\ref{tb:regions}), although with only a few counts the
uncertainty on the flux is high.
Assuming the radio plasma is static, the SSC
predictions for an equipartition magnetic field of 28\,nT
are a factor $\geq 10^2$ lower than the X-ray flux density.
The iC/CMB prediction is closer and
can be brought into agreement with the observations
if the magnetic field is 3$\pm1$\,nT (Figure~\ref{fg:northfaint}). This
field is a factor of $\sim 7-10$ lower than equipartition, a larger
departure than is generally observed \citep{2005ApJ...626..733C}.
Since we do not have observations of
radio emission from the low-energy electrons
responsible for the iC/CMB X-rays, that radio-emitting region may be
more extended (as {\it e.g.} in \citet{2006ApJ...647L.107S}), which
would result in a field closer to the equipartion value. 
Our upcoming, deeper \chandra\ observations (approved in Cycle 13) will
provide improved constraints on these models.

\section{Diffuse, Extended X-ray emission unassociated with radio
  structures.}
\label{sec:cluster}

There is evidence for a cluster
surrounding 3C\,270.1 in optical and infrared data \citep{2009ApJ...695..724H}.
An obvious question, therefore, is whether the diffuse X-rays which
are not associated with the radio structures could be
thermal emission from X-ray gas in the cluster. For the X-rays 
to be cluster emission, we would expect a smooth and
relatively isotropic distribution. The cluster center is not
well-constrained. It was estimated to lie $\sim 20$\arcsec east of 3C\,270.1
by \citet{2009ApJ...695..724H}. 
The current dataset 
has insufficient observed counts (22.8$\pm$5.6, 0.3-8 keV) 
to constrain the spatial distribution of the diffuse emission. However
the existence of an excess of counts around 3C\,270.1, 
and the lack of a similar excess to the east 
suggest that the quasar may lie close to the center of the cluster.

The extended emission has too few counts to constrain the spectral parameters.
The hardness ratio HR=$-0.09\pm0.22$ indicates a
temperature of $\sim 4$ keV in an APEC model but remains within
$2 \sigma$ of a power-law spectrum with $\Gamma \sim 1.7$, so we
cannot rule out non-thermal emission.
An APEC model was fitted, assuming abundance 
Z=0.5 Z$_{\bigodot}$ and
temperature 4 keV at the redshift of the source. With a
grouping of 5 counts per bin to distribute them throughout the band, the 
spectrum looks reasonable, but the signal-to-noise
is too low to provide meaningful constraints.
The fitted normalization yields a broad-band flux,  
F(0.3$-$8\,keV)=$(1.1 \pm 0.5) \times 10^{-14}$ erg cm$^{-2}$
s$^{-1}$ including a correction for the excluded 
sky area assuming isotropic emission. This 
translates to a broad-band X-ray 
luminosity $\sim 2 \times 10^{44}$ erg s$^{-1}$ for
our assumed cosmology, 
consistent with the luminosity of a 4 keV cluster
based on the low-redshift luminosity-temperature (L-T, $<$R$_{500}$)
relation of \citet{2009A&A...498..361P}. 
The few high-redshift ($z > 1.4$) clusters with X-ray measurements
tend to be faint for their temperature based on a self-similar model for their
evolution in comparison with similarly selected, lower-redshift
clusters \citep{2011MNRAS.tmp...73A,2010ApJ...718..133H}.
Our uncertainties are too large to test this for 3C\,270.1. 
Early cluster formation, consistent with a cluster around 3C\,270.1,
would be challenging for cluster formation models,
which suggest that massive dark matter halos form primarily at 
$z{_<\atop^{\sim}}$ 1.2 \citep{2007MNRAS.374.1303C}.

The data are too limited to constrain the X-ray
spatial distribution, temperature, or abundances of the diffuse component. 
Our analysis is based on
$\sim 60$ kpc close to or at the center of the cluster, 
where the physical conditions and the relation to
the full R$_{500}$ luminosity and temperature depend on the 
presence/absence of a cooling core \citep{2009A&A...498..361P}
and can be disturbed by heat input by the quasar.
Deeper \chandra\ X-ray observations are required to confirm the
extended nature and constrain the spatial distribution, spectral
properties, and luminosity of the diffuse X-ray emission.

Qualitatively, the observations of extended X-ray emission are
consistent with the asymmetric depolarization of the radio source
observed by \citet{1991MNRAS.250..171G} (see \S 3.1). 
For a more quantitative
comparison we can consider the simple analysis of
\citet{1991MNRAS.250..198G},
which assumes that the density profile of the cluster can be
modeled as a $\beta$ model and that the energy density in the magnetic
field of the depolarizing medium scales as the energy density in
thermal particles. The observed ratio of the dispersions in
the Faraday depth in
the two lobes, $r_\Delta$ in the notation of Garrington \& Conway, is
$\sim 2$. 
Since we detect an excess of counts close to the quasar, the 
core radius $a$ of any $\beta$ model that
represents the observed X-ray emission must be comparable to or less
than the size of our extraction region, with outer radius
$7.5''$ (64 kpc). From the radio data, the
projected linear size of each lobe is 6\arcsec\ ($\sim 50$ kpc).
The combination of these values and the observed $r_\Delta = 2$
suggest that the gas distribution is rather flat ($\beta \la
0.35$) and that the lobes are not aligned 
close to the line of sight. However
the current quality of X-ray data on the gas near 3C\,270.1 is too low 
to provide useful constraints. 
Our upcoming, deeper \chandra\ observations of this 
source, or {\it Chandra} observations of a
larger sample of objects with observed depolarization, will give us a
probe of the run of gas density and magnetic field strength with radius
in high-$z$ clusters which will be difficult to obtain in any other way.

\section{Conclusions}
\chandra\ X-ray observations of the bright, high-redshift ($z$=1.532)
quasar 3C\,270.1 show strong, unabsorbed power-law emission with a
slope $\Gamma = 1.66 \pm 0.08$, consistent with expectations for
radio-loud quasars. 

Extended X-ray emission associated with the southern radio lobe
of 3C\,270.1 most likely originates in one/both components of the
double hotspot within that
lobe. {\it Spitzer} upper limits for the
hotspot are inconsistent with synchrotron emission from a single power-law
population of electrons as the emission mechanism.
The X-ray emission is consistent with SSC 
for a magnetic field of 24\,nT, about a third of the
equipartition field for that region. 

No X-ray emission is detected from the northern radio hotspot as
expected from our models based on the observed SED. Faint
emission is present just south of the lobe but seems unlikely
to be associated given the lack of alignment. 

Weak but significant
X-ray emission associated with the south-eastern extension of the
northern radio-lobe, coincident with a bend in the radio counter-jet
(``N-C-jet-bend''),
can be explained by iC/CMB emission with a field of 3$\pm1$\,nT, a factor of
$7-10$ lower than the equipartition field. This is a larger departure than is 
typical suggesting that, for example, the emitting region may be larger than
our estimate. Our upcoming, deeper \chandra\ observations will better contrain
the emission in this region.

Extended X-ray emission unassociated with the observed radio
structures may be 
thermal emission from gas within the cluster believed to be
surrounding the quasar or iC emission associated with radio emission
below the present detection limit. 
There are too few counts to constrain its
spatial distribution or spectrum, but assuming
a temperature of $4$ keV yields a luminosity
$\sim 2 \times 10^{44}$ erg s$^{-1}$, consistent with the low-redshift
L-T relation and suggestive of a fully-formed cluster at redshift
1.532. 

Upcoming deeper \chandra\ and radio (EVLA) 
observations will allow us to confirm the location
and origin of the radio-associated X-ray emission for both northern
and southern radio emission structures, to study the spatial distribution
of the diffuse X-ray emission and its relation to any fainter,
diffuse radio emission or to the optical/IR cluster, and to 
constrain its spectral form.

\section*{Acknowledgements}
Support for this work was provided by the National Aeronautics and
Space Administration through \chandra\ Award Number G08-9106X issued by the
\chandra\ X-ray Center, which is operated by the Smithsonian
Astrophysical Observatory for and on behalf of the National
Aeronautics Space Administration under contract NAS8-03060 
(\chandra\ X-ray Center). 
The National Radio Astronomy Observatory is a facility of the National
Science Foundation operated under cooperative agreement by Associated
Universities, Inc..
Observations reported here were obtained at the MMT Observatory, 
a joint facility of the Smithsonian Institution and the University 
of Arizona.
This work is based in part on observations made with the Spitzer
Space Telescope, which is operated by the Jet Propulsion Laboratory,
California Institute of Technology under a contract with NASA.

\bibliography{refs}
\bibliographystyle{apj}

\end{document}